\documentclass[a4paper,11pt]{article}
\pdfoutput=1 

\usepackage{jcappub} 

\usepackage[T1]{fontenc} 
\usepackage{physics}

\title{\boldmath An implementation of nDGP gravity in Pinocchio}

\author[a]{Yanling Song,}
\author[b,c,1]{Bin Hu,\note{Corresponding author}}
\author[d]{Cheng-Zong Ruan,}
\author[e,f,g,h]{Chiara Moretti,}
\author[i,g,h,j]{Pierluigi Monaco}

\affiliation[a]{Center for Gravitation and Cosmology, College of Physical Science and Technology,\\ 
Yangzhou University, Yangzhou 225009, China}
\affiliation[b]{Institute for Frontier in Astronomy and Astrophysics, Beijing Normal University, Beijing, 102206, China}
\affiliation[c]{Department of Astronomy, Beijing Normal University, Beijing, 100875, China}
\affiliation[d]{Institute of Theoretical Astrophysics, University of Oslo, 0315 Oslo, Norway}
\affiliation[e]{SISSA - International School for Advanced Studies, Via Bonomea 265, 34136 Trieste, Italy}
\affiliation[f]{Centro Nazionale ``High Performance Computer, Big Data and Quantum Computing''}
\affiliation[g]{INAF -- Osservatorio Astronomico di Trieste, Via Tiepolo 11, I-34143 - Trieste, Italy}
\affiliation[h]{IFPU -- Institute for Fundamental Physics of the Universe, Via Beirut 2, 34014, Trieste, Italy}
\affiliation[i]{Dipartimento di Fisica dell'Universit\'a di Trieste, Sezione di
Astronomia, via Tiepolo 11, I-34143 Trieste, Italy}
\affiliation[j]{INFN -- Sezione di Trieste}

\emailAdd{bhu@bnu.edu.cn}

\abstract{In this paper we investigate dark matter structure formation in the normal branch of the Dvali-Gabadadze-Porrati (nDGP) model using the PINOCCHIO algorithm. We first present 2nd order Lagrangian perturbation theory for the nDGP model, which shows that the 1st- and 2nd-order growth functions in nDGP are larger than those in $\Lambda$CDM. We then examine the dynamics of ellipsoidal collapse in nDGP, which is accelerated compared to $\Lambda$CDM due to enhanced gravitational interactions. Running the nDGP-PINOCCHIO code with a box size of 512 ${\rm Mpc}~h^{-1}$ and $1024^3$ particles, we analyze the statistical properties of the output halo catalogs, including the halo power spectrum and halo mass function. The calibrated PINOCCHIO halo power spectrum agrees with N-body simulations within $5\%$ in the comoving wavenumber range $k<0.3~(h~{\rm Mpc}^{-1})$ at redshift $z=0$. The agreement is extended to smaller scales for higher redshifts. For the cumulative halo mass function, the agreement between N-body and PINOCCHIO is also within the simulation scatter.}

\begin{document}
\maketitle
\flushbottom



\section{Introduction}

The accelerated expansion of the Universe at late times, as determined from observations of Type Ia supernov\ae \cite{riess1998, perlmutter1999}, has remained an open question for many years. The simplest explanation is given by the cosmological constant, denoted as $\Lambda$, which is included in the standard $\Lambda$CDM cosmological model. The latter provides a good fit to most cosmological observations, however, the increasing precision of recent measurements has brought to light tensions in the cosmological parameters as determined from late versus early time observations. One such tension is the measurement of the Hubble parameter inferred from the Cosmic Microwave Background radiation (CMB) and the one measured from the distance ladder \cite{Freedman:2017yms}, suggesting the possibility of a scenario beyond the standard one. Aside from the issues that emerge at the background level, at the perturbation level the mass variance on a scale of 8~${\rm Mpc}~h^{-1}$ ($\sigma_8$) also shows a tension between CMB lensing and cosmic shear surveys \cite{2020sdljf,2013MNRAS.432.2433H,2021Aerqwrq,2022PhRvD.105b3514A,Dalal:2023olq}. Many alternative theories to $\Lambda$CDM have been proposed to give a more natural interpretation to cosmic acceleration, such as Dark Energy (DE) models and Modified Gravity (MG) models. For DE models, a new dynamical scalar component is added in the stress energy tensor to explain the accelerated expansion; examples of such models include K-essence and Quintessence. The cosmological constant $\Lambda$ can be treated as the minimum of the scalar potential. But the fine-tuning problem between the tiny value of DE invoked by the cosmological accelerated expansion and the Planck scale causes another question. For MG models, modifications to General Relativity (GR) also lead to an additional scalar degree of freedom, which manifest as scalar-tensor theories on the cosmological scales of interest.

The Dvali-Gabadadze-Porrati (DGP) brane world model \cite{Dvali:2000} is a widely studied MG model, which describes our cosmology as a four-dimensional brane embedded in a five-dimensional Minkowski space. On large scales, gravity leaks off the 4-dimensional Minkowski brane into the 5-dimensional “bulk” Minkowski spacetime. On small scales, gravity is effectively bound to the brane and 4-dimensional Newtonian dynamics is recovered to a good approximation.
On scales larger than a cross-over scale $r_c$, gravity behaves as a five-dimensional force. On scales smaller than $r_c$, gravity behaves as a four-dimensional force which can be effectively described by scalar-tensor theories \cite{Nicolis:2004es,Park:2010}. The scalar degree of freedom comes from the displacement of the brane in the bulk, namely brane bending mode. 

Observations in the Solar System require that any modified gravity model has to restore to GR on these scales, while the explanation late-time acceleration of the universe under the frame of modified gravity requires deviations of GR. 
In order to satisfy both small- and large-scale requirements, the extra fifth gravitational force has to be shielded in high density regions. This is achieved by means of a so-called screening mechanism. In the DGP model this is achieved by the Vainshtein screening \cite{Babichev:2013usa}. Vainshtein screening relies on the nonlinear interactions of the second derivatives of the scalar field. When the local curvature is large $|\nabla^2\Psi_{\rm N}|>\Lambda_{\rm T}^3$, the screening mechanism becomes active. 

There are two branches of the DGP model, the self-accelerated branch (sDGP) and the normal branch (nDGP), corresponding to the two possible ways of embedding the brane in the bulk. The sDGP branch is able to achieve acceleration in the late time Universe, but its expansion history substantially conflicts with observational data \cite{Fairbairn:2005ue}. Besides, its perturbation in the de Sitter background is plagued by a ghost instability \cite{Luty:2003,Koyama:2007za}. On the other hand, nDGP does not suffer from ghost instability issues, although it does not lead to self-acceleration. One needs to add an extra DE component in the stress energy tensor to account for the accelerated expansion. The DE component can be freely tuned, so that the expansion history matches $\Lambda$CDM \cite{Sahni:2003es,Schmidt:2009b}~\footnote{This is an assumption made for ease of comparisons to $\Lambda$CDM simulations, and is not a strict observational requirement \cite{Lombriser:2009}.}. This model, called nDGP+DE, allows for a separation of the MG effect from the background geometry and purely study its effect on the linear and non-linear scales of structure formation. 

The Large Scale Structure of the Universe is believed to arise from primordial matter density fluctuations subject to gravitational instability, and is therefore a powerful probe of gravity models. The high-quality observational data that will be provided by upcoming and ongoing Stage IV galaxy surveys  are expected to be able to distinguish between different gravity models. Such a task can only be achieved with the development and validation of robust analysis pipelines, which usually include the generation of high-quality, realistic simulated data for all the alternative models. In particular, large numbers of simulations are needed for the computation of numerical covariance matrices for cosmological observables.
At present, the most precise numerical tool to generate the mock data are N-body simulations (see \cite{angulo2022} for a review). 
For the nDGP model, there exist several N-body softwares, including \cite{Khoury:2009tk}, DGPM \cite{Schmidt:2009a}, \textsc{ecosmog} \cite{Li:2011vk,Li:2013nua,Barreira:2015xvp}, \textsc{MG-Arepo} \citep{Hern.2021MNRAS.503.3867H.MGGLAM} and \textsc{MG-GLAM} \cite{Hernandez-Aguayo:2021kuh,Ruan:2021wup}. 
However, the generation of a large number of large volume, high resolution N-body simulations can be computationally prohibitive, especially for MG models. 
The PINOCCHIO algorithm \cite{Monaco:1997cq,Monaco:2001jg,Monaco:2001jf,Taffoni:2001jh,Monaco:2013qta} (PINpointing Orbit-Crossing Collapsed Hierarchical Objects) is a semi-analytical software to generate halo catalogs, able to run on personal laptops at a fraction of the time needed by a full N-body simulation. The code was extended to massive neutrino cosmologies in \cite{Rizzo:2016mdr}, $f(R)$ modified gravity in \cite{moretti2020} and the Cubic Galileon model in \cite{Song:2021msd}. Here, we further extend it to the nDGP model, describing suitable extensions both of Lagrangian Perturbation theory and ellipsoidal collapse, on which the code relies. We then study the linear and mildly-nonlinear regimes of structure formation and compare the PINOCCHIO output to N-body simulations. Another fast approximate simulation algorithm, which is a hybrid of Lagrangian perturbation theory with a particle-mesh solver, is the COLA (COmoving Lagrangian Acceleration) package \cite{Tassev:2013pn, winther2017}. Very recently, a matter power spectrum emulator based on this method has been successfully coded for nDGP model \cite{Fiorini:2023fjl}. 

This paper is organized as follows: in Section~\ref{sec:model} we describe the background, linear and nonlinear perturbation evolution of the nDGP model.
In Section~\ref{sec:3}, we briefly introduce the Lagrangian perturbation theory and apply it to nDGP to calculate the 1st- and 2nd- order growth functions of galaxy clustering. In Section~\ref{sec:4}, we describe the prescription implemented in the “nDGP-PINOCCHIO” algorithm, including the extension of ellipsoidal collapse. In Section~\ref{sec:5}, we show the halo power spectrum and mass function obtained with the code, comparing with those obtained from N-body simulations. Finally, we present our conclusions in Section~\ref{sec:6}.

\section{nDGP model}
\label{sec:model}

The action of the DGP model is composed by a four- and a five- dimensional part \cite{Dvali:2000}
\begin{equation}
S_{\rm DGP}=\int d^5x\sqrt{-g^{(5)}} \frac{R^{(5)}}{16\pi G^{(5)}} + \int d^4x\sqrt{-g}\Big( \frac{R}{16\pi G} + L_m \Big) \, ,  
\end{equation}
where $g^{(5)},R^{(5)},G^{(5)}$ are respectively the metric determinant, the Ricci scalar and the gravitational constant in the five-dimensional bulk, and $g,R,G$ are corresponding quantities in our four-dimensional universe. $L_m$ is the Lagrange density for all matter fields. $G^{(5)},G$ can be used to define the cross-over scale, which is the only additional parameter in DGP model:
\begin{equation}
r_c=\frac{1}{2}\frac{G^{(5)}}{G} \, . 
\end{equation}
Below this cross-over scale gravity looks four-dimensional, above it gravity needs to be described in a five-dimensional way. As mentioned in the introduction, there are two branches of solutions in a cosmological background: the self-accelerating branch (sDGP) and the normal branch (nDGP).

In the spatially flat Friedmann-Lemaître-Robertson-Walker (FLRW) metric, the background evolution of the universe can be written as \cite{Deffayet:2001his,Deffayet:2002}:
\begin{equation}
H(a)=H_0\sqrt{\Omega_{m0}~a^{-3}+\Omega_{rc}}\pm \sqrt{\Omega_{rc}} \, , 
\end{equation}
where $H_0$ is the Hubble parameter evaluated at present time, $\Omega_{m0}=8\pi G\bar\rho_{m0}/(3H_0^2)$ is the present day fractional matter density and $\Omega_{rc}=1/(4H_0^2r_c^2)$, and $a$ is the scale factor. The sDGP branch corresponds to the $-$ sign, while the nDGP branch corresponds to the $+$ sign. In this paper we study the stable normal branch nDGP, in which an extra DE component is added and adjusted so that it mimics the $\Lambda$CDM background evolution.  When $r_c\rightarrow \infty$, the nDGP model reduces to $\Lambda$CDM.

For the perturbed FLRW metric in the conformal Newtonian gauge
\begin{equation}
ds^2=(1+2\Psi)dt^2-a^2(t)(1-2\Phi)d\vec x^2 \, , 
\end{equation}
on scales much smaller than both the horizon and the cross-over scale one can apply the quasi-static approximation, i.e. one can neglect time-derivatives \cite{Koyama:2007es,Schmidt:2009a,Winther:2015es}. One can then write the evolution equations for the potentials as:
\begin{equation}
\nabla^2\Psi=4\pi Ga^2\bar\rho_m\delta+\frac{1}{2}\nabla^2\phi,
\label{eq:poisson}
\end{equation}
\begin{equation}
\nabla^2\phi + \frac{r_c^2}{3\beta a^2}\Big[(\nabla^2\phi)^2 - (\nabla_i\nabla_j\phi)(\nabla^i\nabla^j\phi)  \Big]= \frac{8\pi G}{3\beta}a^2\bar\rho_m\delta\;,
\label{eq:scalar_field}
\end{equation}
where $\phi$ is a scalar degree of freedom corresponding to the brane bending mode, $\delta=(\rho_m-\bar\rho_m)/\bar\rho_m$ is the density contrast, $\nabla$ is the comoving spatial derivative, and 
\begin{equation}
\beta(a)=1+ 2Hr_c\Big( 1+\frac{\dot H}{3H^2} \Big)=1+\frac{\Omega_{m0}~a^{-3}+2\Omega_{\Lambda 0}}{2\sqrt{\Omega_{rc}~(\Omega_{m0}~a^{-3}+\Omega_{\Lambda 0})}} \, .
\end{equation}
Combining Eqs.~(\ref{eq:poisson})~(\ref{eq:scalar_field}), one can get the modified Poisson equation for linear perturbations
\begin{equation}
\nabla^2\Psi=4\pi G_{\rm lin}a^2\bar\rho_m\delta=4\pi Ga^2 \left(1+\frac{1}{3\beta} \right)\bar\rho_m\delta \,,
\label{eq:linear_poisson}
\end{equation}
where $G_{\rm lin}$ represents the effective gravitational constant on linear scales and is defined as $G_{\rm lin}=G \qty(1+\frac{1}{3\beta})$. On scales where the local density is much larger than the average density, the linear description of Eq.~(\ref{eq:linear_poisson}) is not valid. Additionally, the effective gravitational constant must recover the GR value on such small scales: for the DGP model, this is achieved by nonlinear kinetic interactions, namely the Vainshtein screening mechanism \cite{vainshtein1972, Babichev:2013usa}. When considering second order perturbations, solving Eqs.~(\ref{eq:poisson})~(\ref{eq:scalar_field}) for a top hat overdensity profile, one can get \cite{Koyama:2007es,Schmidt:2010}
\begin{equation}
\nabla^2\Psi=4\pi G_{\rm NL}a^2\bar\rho_m\delta=4\pi Ga^2\qty[1+\frac{2}{3\beta}\frac{r^3}{r_V^3}\qty(\sqrt{1+\frac{r_V^3}{r^3}}-1) ]\bar\rho_m\delta\;,
\end{equation}
where 
\begin{equation}
r_V^3=\frac{16r_c^2G\delta M}{9\beta^2}
\end{equation}
is called the Vainshtein radius and $\delta M=\frac{4}{3}\pi r^3\bar\rho_m\delta$ is the mass enclosed in a sphere of radius $r$. $G_{\rm NL}$ represents an effective gravitational constant which encodes the effect of MG on mildly nonlinear scales. One can see that on scales much smaller than the Vainshtein radius $r\ll r_V$, the effective gravitational constant recovers the value of GR. In contrast with the chameleon screening mechanism, Vainshtein screening shields the fifth force via the high curvature of the local object. Outside the local object, the environment density need not to be very high for Vainshtein screening to be effective, which is a requirement for chameleon screening.

\section{Lagrangian perturbation theory for nDGP}\label{sec:3}

Large scale structure is believed to arise from small initial perturbations. The corresponding dynamics can be described by standard Eulerian perturbation theory (SPT) or by Lagrangian perturbation theory (LPT) (see \cite{Bouchet:1996ez} for a review).  SPT is described in the Eulerian frame, and the density contrast and peculiar velocity are expanded in Euclidean coordinates as
\begin{eqnarray}
    \delta =\varepsilon \delta^{(1)}+\varepsilon^{2}\delta^{(2)}+..., \\
    \theta =\varepsilon \theta^{(1)}+\varepsilon^{2}\theta^{(2)}+...,
\end{eqnarray}
where $\theta=\nabla\cdot v$ is the divergence of the peculiar velocity and one can ignore the vorticity of the velocity which decays due to the expanding universe. $\varepsilon$ is a small quantity, which is factorized to label the perturbation order. Instead of working in the Euclidean coordinate, LPT traces the fluid elements (or particles) in Lagrangian coordinates. The famous Zel'dovich approximation is the first order solution in LPT; it is valid until orbit crossing occurs. The main quantity in LPT is the displacement field $\vec S$, describing the change from the initial position $\vec q$ to the final comoving position $\vec x$:
\begin{equation}
\vec{x}(t,\vec{q})=\vec{q}+\vec{S}(t,\vec{q}) \, .
\label{eq:displ}
\end{equation}
The displacement $\vec S$ can be expanded as
\begin{equation}
\vec{S}=\varepsilon\vec{S}^{(1)}+\varepsilon^{2} \vec{S}^{(2)}+... \, .
\label{eq:expand displ}
\end{equation}
By enforcing mass conservation during evolution $\rho_md^3x=\bar\rho_md^3q$, one can express the displacement field in terms of the density contrast:
\begin{equation}
\frac{1}{J(t,\vec{q})}=1+\delta(t,\vec x) \, ,
\label{eq:delta Jacob}
\end{equation}
where $J(t,\vec q)$ is the determinant of the Jacobian matrix:
\begin{equation}
J_{ij}=\frac{\partial x^{i}}{\partial q^{j}}=\delta_{ij}+\frac{\partial S^{i}}{\partial q^{j}}=\delta_{ij}+S_{i,j} \, .
\label{eq:Jacob}
\end{equation}
Here commas denote differentiation with respect to $\vec q$. 
In order to extend LPT in the context of the particular MG model we are considering, in what follows we derive the equation of motion for particle trajectories in nDGP:
\begin{equation}
\frac{d^{2}\vec{x}}{dt^{2}}+2H\frac{d\vec{x}}{dt}=-\frac{1}{a^{2}}\nabla\Psi.
\label{eq:trajectory}
\end{equation}
Taking the divergence of Eq.~(\ref{eq:trajectory}), then submitting Eq.~(\ref{eq:displ}) to the left hand and Eq.~(\ref{eq:poisson}), (\ref{eq:scalar_field}) to the right hand, we obtain:
\begin{equation}
\nabla\cdot\hat T\vec S=-A(a)\delta-B(a)\Big[ \qty(\frac{\nabla^2\phi}{a^2})^2 - \qty(\frac{\nabla_i\nabla_j\phi}{a^2})^2 \Big] \, ,
\label{eq:main}
\end{equation}
where we used the following definitions:
\begin{eqnarray}
\hat T&=&\frac{d^{2}}{dt^{2}}+2H\frac{d}{dt} \, , \\
A(a)&=&4\pi G\bar\rho_m(1+\frac{1}{3\beta}) \, , \\
B(a)&=&-\frac{r_c^2}{6\beta} \, . 
\end{eqnarray}
Eq.~(\ref{eq:main}) is expressed in Eulerian coordinates. To transform this equation into Lagrangian space, we need to transform the spatial derivative operator to the one in Lagrangian space. The coordinate transformation can be achieved by the Jacobian matrix:
\begin{equation}
\nabla_{i}=(J^{-1})_{ij}\nabla_{qj} \, ,
\end{equation}
where $\nabla_{qj}$ is the $j$ component spatial derivative in Lagrangian space. Besides,  $\phi$ can be replaced by the density field using Eq.~(\ref{eq:scalar_field}), and $\delta=\frac{1-J}{J}$. 
All the ingredients of the Jacobian matrix are directly related to the displacement field through Eq.~(\ref{eq:Jacob}). 
For this reason, the Jacobian can be expanded into different orders of perturbations\footnote{Interested readers can look into \cite{Aviles:2017aor} for more details about the expanding formulas of the Jacobian matrix.}. 
After these operations, we are able to solve Eq.~(\ref{eq:main}) and get the solutions for the first and second order displacements $\vec S^{(1)},\vec S^{(2)}$.  
The detailed steps are very similar to those in the Cubic Galileon model described in detail in \cite{Song:2021msd}.  
For instance, Eq.~(\ref{eq:main}) is the same as Eq.(27) in \cite{Song:2021msd}, just with different functional forms for $A(a),B(a)$. 
Here, we directly show the results in Fourier space. At the first order
\begin{equation}
S_{i,i}^{(1)}(\vec k,t)=-D_1(t)\delta^{(1)}(\vec k,t_0) \, ,
\label{eq:displ1}
\end{equation}
where the linear growth factor $D_1(t)$ satisfies 
\begin{equation}
(\hat T-A(a))D_1(t)=0 \, . 
\label{eq:D1}
\end{equation}
One can see that $D_1$ is dependent on $t$ only. This is because in models that feature the Vainshtein screening mechanism the scalar field is effectively massless. As a consequence the Compton wavelength scale, above which deviations from $\Lambda$CDM emerge, is beyond the Hubble horizon. The linear gravitational constant $G_{\rm lin}$ is therefore scale-independent, which results in $D_1$ being scale independent as well.

For second order perturbations, we have
\begin{eqnarray}
S_{i,i}^{(2)}(\vec k,t)=-\frac{1}{2}D_2(t)\int\frac{d^3\vec k_1d^3\vec k_2}{(2\pi)^3}\delta(\vec k-\vec k_1-\vec k_2) 
\qty[1-\frac{(\vec k_1\cdot\vec k_2)^2}{k_1^2 k_2^2}]\delta_{k_1}^{(1)}(t_0)\delta_{k_2}^{(1)}(t_0) \, ,
\label{eq:displ2}
\end{eqnarray}
where $D_2(t)$ satisfies 
\begin{equation}
[\hat T-A(a)]D_{2}(t)=2C(a)D_1^2(t),
\label{eq:D2}
\end{equation}
with
\begin{equation}
C(a)=\frac{1}{2}A(a)+B(a) \qty(\frac{8\pi G\bar\rho_m}{3\beta})^2. 
\end{equation}
$D_2$ is also independent of $k$ , because we absorb the $\left[1-\frac{(\vec k_1\cdot\vec k_2)^2}{k_1^2 k_2^2}\right]$ term into the initial convolution of the displacement field, leaving only the time-dependent term. 
Changing the time parameter from physical time $t$ to the scale factor $a$, we have
\begin{equation}
\hat T=\frac{d^{2}}{dt^{2}}+2H\frac{d}{dt}=a^{2}H^{2}\left[\frac{d^{2}}{da^{2}}+ \qty(\frac{H^{'}}{H}+\frac{3}{a})\frac{d}{da}\right] \, , 
\end{equation}
where $\prime$ represents derivative with respect to $a$. We use a fourth-order Runge-Kutta integration method to solve the ordinary differential Eqs.~(\ref{eq:D1}) -- (\ref{eq:D2}), with initial conditions $D_1(a)=a,\frac{dD_1}{da}=1$ for $D_1$, $D_2(a)=\frac{3}{7}a^2,\frac{dD_2}{da}=\frac{6}{7}a$ for $D_2$.

\section{Implementation of nDGP in PINOCCHIO}\label{sec:4}

PINOCCHIO
is a semi-analytical algorithm for generating realisations of dark matter halos in cosmological volumes, including their hierarchical formation histories. It is based on excursion set theory, using the dynamics of LPT and ellipsoidal collapse. The code begins by creating a realisation of a linear density field sampled on a grid. The evolution of particles (or grid points) is described through LPT, and their position can be computed for any given time. According to Eq.~(\ref{eq:delta Jacob}), $\delta$ approaches infinity when $J\rightarrow 0$: this is when particles reach orbit crossing and form multi-stream regions, thus creating caustics. Hence, $J\rightarrow 0$ is used to define the collapse time of fluid elements (that can be thought as particles in the corresponding N-body simulation).  A collapsed particle is marked as halo, and then neighboring particles are grouped together by a fragmentation algorithm which determines membership to halos or filaments.

The determinant of the Jacobian matrix reads
\begin{equation}
J=\det(\delta_{ij}+S_{i,j})=(1-\lambda_{1}(a))(1-\lambda_{2}(a))(1-\lambda_{3}(a)) \, ,
\label{eq:jacobian}
\end{equation}
where $\lambda_1>\lambda_2>\lambda_3$  are the eigenvalues of $-S_{i,j}(a,\vec q)$. Collapse can thus take place along the three directions of the eigenvector of $-S_{i,j}(a,\vec q)$. 
In PINOCCHIO, the collapse time for a particle is defined as the moment when $\lambda_1\rightarrow 1$, namely when the first axis collapses. 

By Taylor-expanding the gravitational potential to second order, it is easy to show that the evolution of the mass element is analogous to the evolution of a homogeneous ellipsoid that starts as a sphere and is distorted by the tidal field. Collapse time can then be computed by using the formalism of ellipsoidal collapse.
Following and extending the approach of \cite{Bond:1996es}, we can derive the evolution for the principal axes of the ellipsoid $a_i$ ($i=1,2,3$) for the nDGP model:
\begin{equation}
\frac{d^2a_i}{dt^2}=\frac{\ddot a}{a}a_i-4\pi G_{\rm NL}\bar\rho_m \qty[\frac{\delta}{3}+\frac{5}{4}b_i(t)+\frac{b_i(t)}{2}\delta].
\label{eq:ellip}
\end{equation}
where $\ddot a$ is the second order time derivative of the scale factor, and 
\begin{equation}
\quad b_{i}(t)=-\frac{2}{3}+a_{1}(t)a_{2}(t)a_{3}(t)\int^{\infty}_{0}\frac{d\tau}{[a_{i}^{2}(t)+\tau]\Pi^{3}_{j=1}(a_{j}^{2}(t)+\tau)^{1/2}}.
\label{eq:int}
\end{equation}
The deviation from $\Lambda$CDM is encoded in the $G_{\rm NL}$ function discussed in the previous sections. Eq.~(\ref{eq:ellip}) is an integro-differential equation, which should in principle be solved for each particle, resulting in a computationally expensive procedure. For our purposes, we adopt the description of ellipsoid evolution presented in \cite{Nadkarni-Ghosh:2014bwa}, that is equivalent to the approach of \cite{Bond:1996es}, but avoids to compute the integral of eq.~(\ref{eq:int}) by recasting the problem into a system of nine ordinary differential equations. To this aim we define nine dimensionless parameters:
\begin{eqnarray}
    \lambda_{\rm a,i}&=&1-\frac{a_{\rm i}}{a}\;, \\
    \lambda_{\rm v,i}&=&\frac{1}{H}\frac{\dot a_{\rm i}}{a_{\rm i}}-1\;,\\
    \lambda_{\rm d,i}&=&\frac{\delta}{3}+\frac{5}{4}b_i^\prime(t)+\frac{b_i^\prime(t)}{2}\delta.
\end{eqnarray}
Here \(\lambda_{\rm a,i}\) correspond to the eigenvalues of the deformation tensor and characterize the shape of the ellipsoid, \(\lambda_{\rm v,i}\) describe the deviation of the velocity of the \(i\)th axis from the background Hubble flow and \(\lambda_{\rm d,i}\) correspond to the eigenvalues of the tensor of
second derivatives of the gravitational potential. Taking the time
derivative of \(\lambda_{\rm a,i},\lambda_{\rm v,i},\lambda_{\rm d,i}\), we obtain this set of first-order ordinary differential equations
\begin{eqnarray}
\label{eq:NGS}
    \frac{d\lambda_{\rm a,i}}{d\ln a}&=&-\lambda_{\rm v,i}(1-\lambda_{\rm a,i})\;,\nonumber\\
    \frac{d\ln\lambda_{\rm v,i}}{d\ln a}&=&-\frac{3}{2}\mu\Omega_{\rm m}\lambda_{\rm d,i}-\lambda_{\rm v,i}(2+\frac{\dot H}{H^2})-\lambda_{\rm v,i}^2\;, \nonumber\\
    \frac{d\lambda_{\rm d,i}}{d\ln a}&=&-(1+\delta)(\delta+\frac{5}{2})^{-1}(\lambda_{\rm d,i}+\frac{5}{6})\sum_{j=1}^{3}\lambda_{\rm v,j}\nonumber\\
    &&+(\lambda_{\rm d,i}+\frac{5}{6})\sum_{j=1}^{3}(1+\lambda_{\rm v,j})-(\delta+\frac{5}{2})(1+\lambda_{\rm v,i})\\
    &&+\sum_{j\neq i}\frac{[\lambda_{\rm d,j}-\lambda_{\rm d,i}][(1-\lambda_{\rm a,i})^{2}(1+\lambda_{\rm v,i})-(1-\lambda_{\rm a,j})^{2}(1+\lambda_{\rm v,j})]}{(1-\lambda_{\rm a,i})^{2}-(1-\lambda_{\rm a,j})^{2}}\;,\nonumber\\
    \mu&=&4\pi a^2\qty[1+\frac{2}{3\beta}\frac{r^3}{r_V^3}\qty(\sqrt{1+\frac{r_V^3}{r^3}}-1) ]\bar\rho_m\delta\;,
\end{eqnarray}
where $\delta=\lambda_{\rm d,1}+\lambda_{\rm d,2}+\lambda_{\rm d,3}$ and $\mu$ is the dimensionless gravitational coupling strength.
The initial conditions for this set of equations are $\lambda_{a,i}=\lambda_{v,i}=\lambda_{d,i}=\lambda_{i}$,
where $\lambda_i$ is the eigenvalue of $-S^{(1)}_{i,j}(\vec k,a_0)$ of Eq.~\ref{eq:jacobian}, since the Zel’dovich approximation is accurate enough at the initial time.
By solving Eq.~(\ref{eq:NGS}) numerically until the first axis reaches $\lambda_{a,i}=1$ we determine the collapse time as the time at which orbit crossing happens.
To improve computing efficiency, these equations are not solved for each particle, but solutions are first computed on a grid of values of $\lambda_i$ to create a look-up table, and collapse times of single particles are obtained by interpolation.
Since PINOCCHIO is, by definition, a perturbative method for clustering of large-scale structures, the screening mechanism, which asks for the high non-linearities, is not directly coded in the clustering processes. It mainly enters through the particle collapsing process in PINOCCHIO algorithm, where the high non-linearity occurs. One can see that the modified gravitational coupling, $\mu$ function, directly enters into the collapsing equations~(\ref{eq:NGS}). Physically, it will slow down the collapsing speed which is predicted by the linear modified gravity effect.      

In the spirit of excursion set theory, the computation of collapse time is applied to a smoothed version of the density field, and repeated several times (of order $\sim10$) for each smoothing scale considered, then the earliest collapse time for each particle is stored. The grouping of particles into halos is performed with an algorithm that mimics their hierarchical clustering, separating particles that belong to halos from those that have undergone orbit crossing but still lie in the filamentary network. 
Particle positions are finally obtained using the displacement field of Eqs.~(\ref{eq:displ1}) -- (\ref{eq:displ2}).
Halos can be output both at fixed time in a periodic box, or on the past light cone, a feature that we do not use in this paper. Please see \cite{Munari:2016aut} and \cite{Song:2021msd} for a more detailed description of the code.

While PINOCCHIO is tailored at the production of halo catalogs, it is possible to use it to generate shapshots of the matter field. This is done by moving particles to their final position according to their state; particles outside halos are moved to their final position using LPT, while particles in halos are distributed around the halo center of mass as spheres, following the NFW (\cite{Navarro:1995iw}) density profile and a Maxwellian velocity profile. This allows us to compute also the matter power spectrum. However, the latter is known to be less accurate than the halo power spectrum. In particular, the level of agreement at high wavenumbers mostly depends on how good the calibration of the mass-concentration relation is, since the PINOCCHIO algorithm is not predictive for the 1-halo term. In what follows we focus therefore on halo statistics such as the halo power spectrum and hao mass function.


\section{Simulations and Results}\label{sec:5}

In this section we validate our implementation for nDGP gravity, and determine the range of vailidity of the PINOCCHIO approximated method. In order to do so, we compare summary statistics computed from PINOCCHIO against those extracted from N-body simulations, focusing on the matter and halo power spectrum in real space and the halo mass function.
We extend the standard PINOCCHIO algorithm to the nDGP model by incorporating the effect of non-standard gravity, as described in the previous sections. We run the code for a box with side L=512 ${\rm Mpc}~h^{-1}$ with 1024$^3$ particles and the same cosmological parameter settings both for PINOCCHIO and the N-body simulation. The initial power spectra are given by the fitting formula of \cite{1998ApJ...496..605E}. The cosmological parameters we adopt are $\Omega_{m0}=0.269, \Omega_{\Lambda0}=0.731, h=0.704,\sigma_8=0.8,n_s=0.966$ for both nDGP and $\Lambda$CDM. For PINOCCHIO, we compute the displacement field of dark matter particles with 2LPT as presented in Section \ref{sec:3}. We run 15 realisations for PINOCCHIO and 5 realisations for N-body simulations, different realisations have different random seeds for the initial conditions. For each realisation, nDGP and $\Lambda$CDM share the same random seed. The fragmentation process that constructs halos in PINOCCHIO is calibrated in the context of $\Lambda$CDM to reproduce the mass function of N-body simulations with halos identified with a friends-of-friends (FoF) algorithm with linking length 0.2 times the mean inter particle distance. We do not perform re-calibration of the fragmentation here.

The $N$-body simulations are run with \textsc{GLAM} \cite{Klypin:2018MNRAS.478.4602K} for $\Lambda$CDM and its modified gravity extension, \textsc{MG-GLAM} \cite{Hernandez-Aguayo:2021kuh}, for the nDGP model.
GLAM is a parallel particle-mesh code for the massive production of N-body simulations and mock galaxy catalogues. 
MG-GLAM uses a multi-grid relaxation technique to solve the non-linear field equations of modified gravity models.
We study two nDGP models with different values of the $r_c$ parameter, $H_0 r_c = 1$ and $5$, along with their $\Lambda$CDM counterparts.
The initial conditions are generated by GLAM using the Zel'dovich approximation at $z_{\mathrm{init}} = 99$, using as input the same linear power spectra as in PINOCCHIO. 

In the N-body simulations, nonlinear matter power spectra and halo catalogues at output redshifts are measured on the fly. 
GLAM employs the Bound Density Maximum (BDM; \cite{Klypin:1997astro.ph.12217K.BDM,Riebe:2011arXiv1109.0003R}) algorithm to identify halos, which is different from the FoF algorithm that PINOCCHIO is calibrated on.
The halo mass definition adopted is 
\begin{align}
    M_{\mathrm{vir}} = \frac{4\pi}{3} R_{\mathrm{vir}}^3 \Delta_{\mathrm{vir}}(z)\rho_{\mathrm{crit}}(z) \, ,
\end{align}
where $\rho_{\mathrm{crit}}(z) \equiv 3H^2(z) / (8 \pi G)$ is the critical density and $\Delta_{\mathrm{vir}}$ is the virial overdensity within the halo radius $R_{\mathrm{vir}}$.
We use the virial density definition for $\Delta_{\mathrm{vir}}$ given by \cite{1998ApJ...495...80B}, 
\begin{align}
    \Delta_{\mathrm{vir}}(z) = 18 \pi^2 + 82 \big[\Omega_{m}(z) - 1\big] - 39 \big[ \Omega_m(z) - 1 \big]^2 \,.
\end{align}

\begin{figure}
	\centering
	\includegraphics[width=0.8\columnwidth]{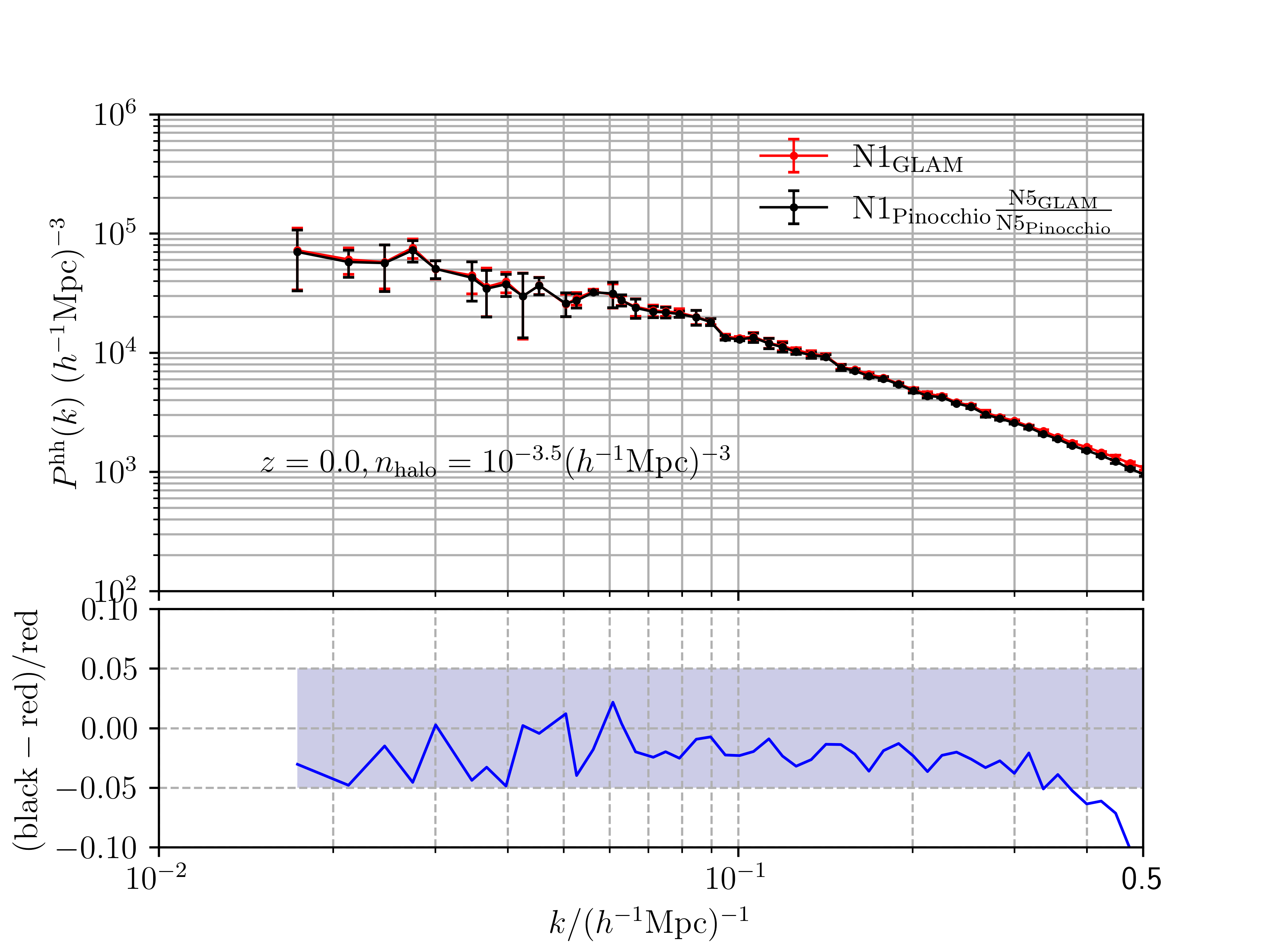} 
    \includegraphics[width=0.8\columnwidth]{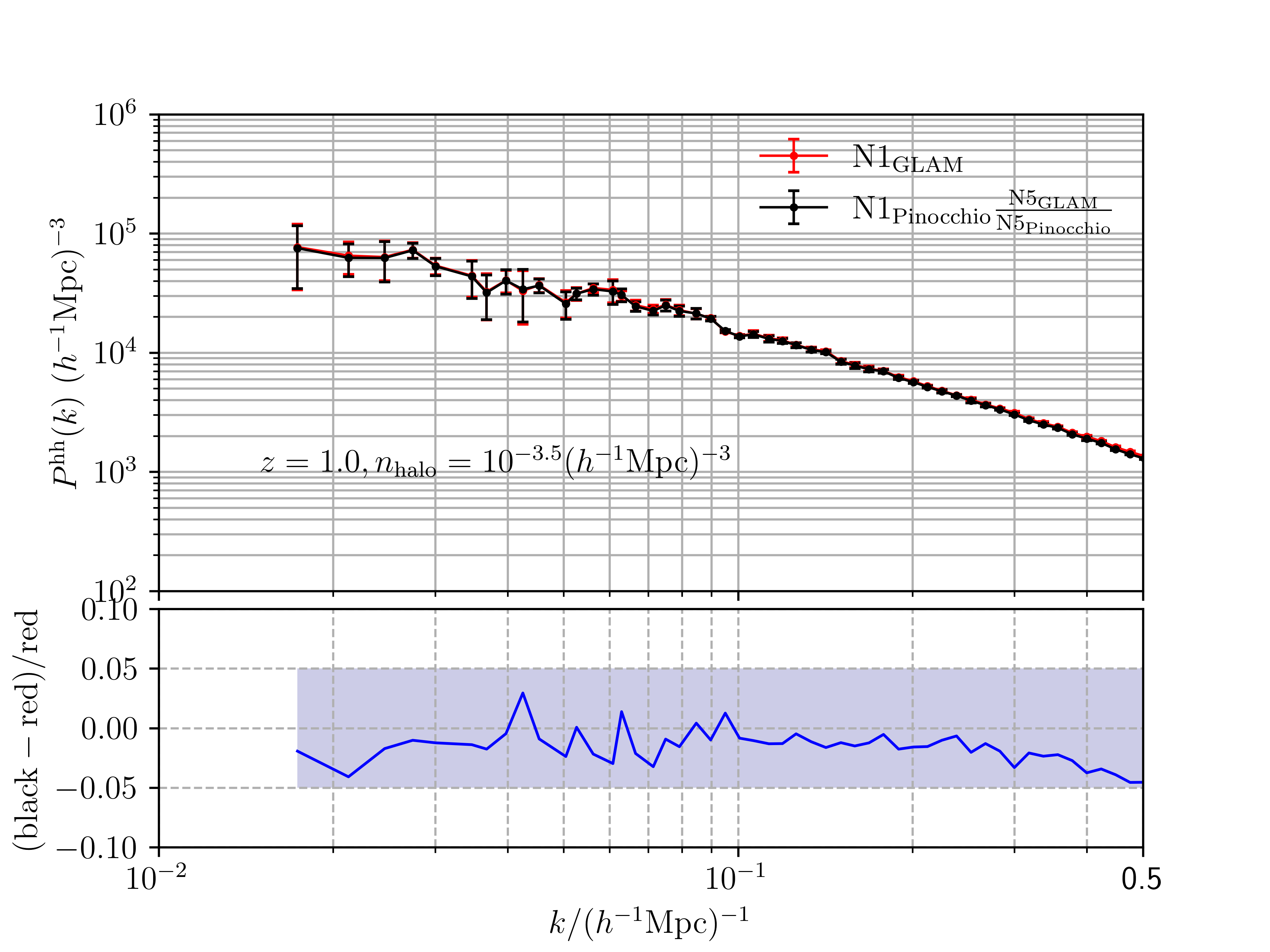} 
    \caption{Halo power spectrum (top panels) and relative residuals (bottom panels) between N-body and rescaled PINOCCHIO for N1 model. The upper and lower plots show the results for $z=0$ and $z=1$, respectively. Red points and error bars are the averaged value and the standard deviations from on 5 N-body simulations, black points and error bars are from 15 PINOCCHIO runs. }
    \label{fig:pkhh}
\end{figure}

The halo catalogs we use to calculate the halo power spectra have the same volume sizes and we filter them to have the same halo number density $n_{\rm halo}=10^{-3.5}~(h^{-1}{\rm Mpc})^{-3}$. 
The reason we match the number density instead of using a minimum halo mass is to mitigate differences in the summary statistics that can be introduced because of the halo finder used in MG-GLAM. 
This is however not sufficient to completely remove the effect, and results in a systematic off-set in the halo power spectra between N-body and PINOCCHIO. We stress that this is not a failure in the approximations implemented in PINOCCHIO, nor of the modified gravity implementation presented here, but solely related to the different halo finder algorithm.
As shown by \cite{Munari:2016aut}, even with an ideal reconstruction of halo number density, PINOCCHIO is able to recover the linear bias $b_1$ of the halo power spectrum within an accuracy of a few percent, the difference being due to the non-perfect reconstruction of halos; this results in a difference in the normalization of halo power spectra of several percent when compared with a simulation. 
To further mitigate this discrepancy, we use the ratio between N-body and N5-PINOCCHIO (smaller deviation from $\Lambda$CDM) to re-scale the N1-PINOCCHIO. We compare this re-scaled PINOCCHIO result with the one from N-body in  Fig.~\ref{fig:pkhh}. The top and bottom panels show the results for $z=0$ and $z=1$ respectively, with red points marking the average from 5 N-body simulations and black points marking the average of 15 PINOCCHIO runs.
In each plot the upper panel shows the halo power spectrum and the lower panel shows the relative residuals between N-body and re-scaled PINOCCHIO for N1 model. We can see that after this correction, PINOCCHIO agrees with N-body simulation within $5\%$ range up to $k=0.3~(h~{\rm Mpc}^{-1})$ for $z=0$ case and  $k=0.5~(h~{\rm Mpc}^{-1})$ for $z=1$ case. This is in line with results obtained for $\Lambda$CDM.

\begin{figure}
	\centering
	\includegraphics[width=0.8\columnwidth]{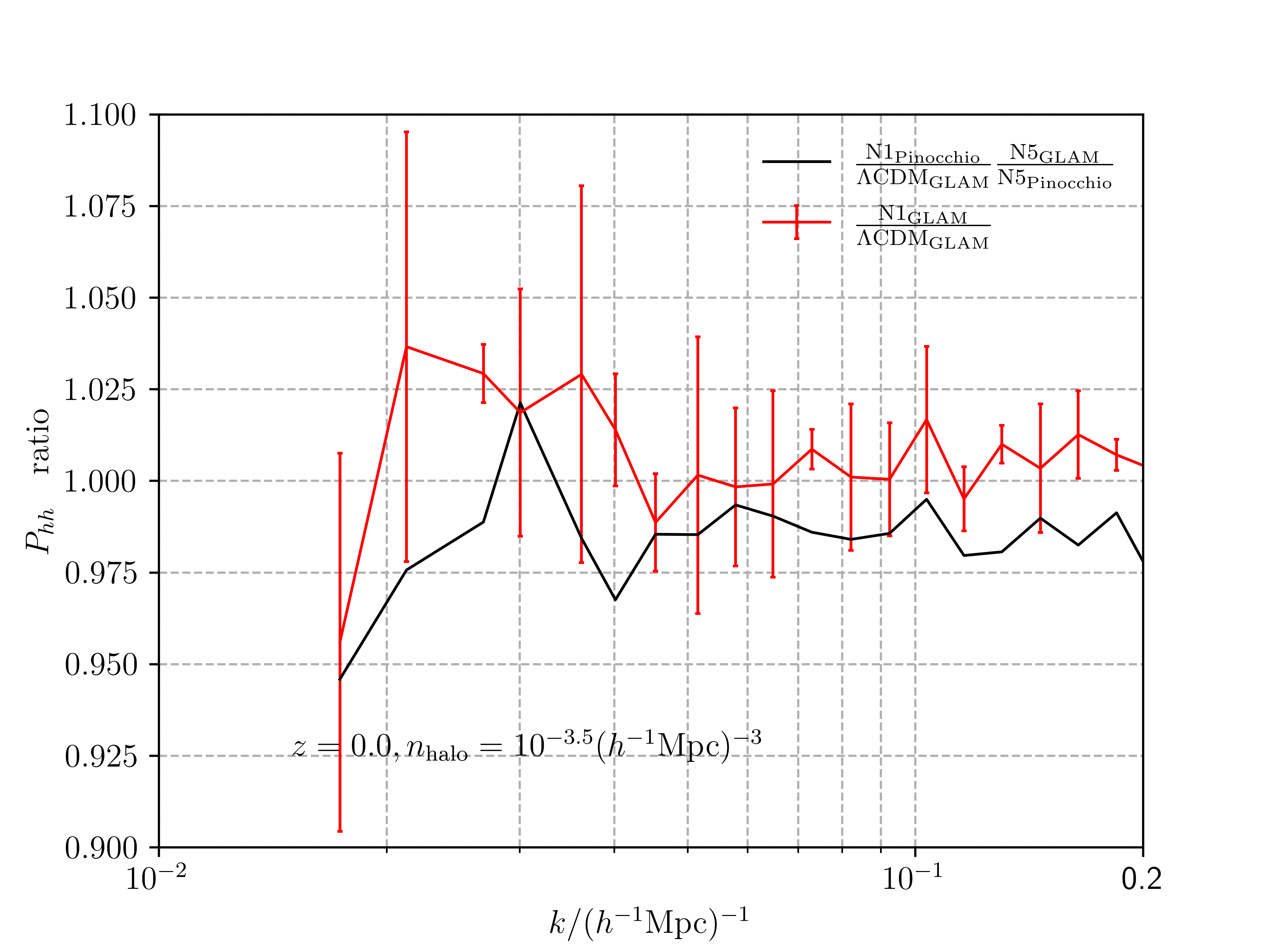} 
    \includegraphics[width=0.8\columnwidth]{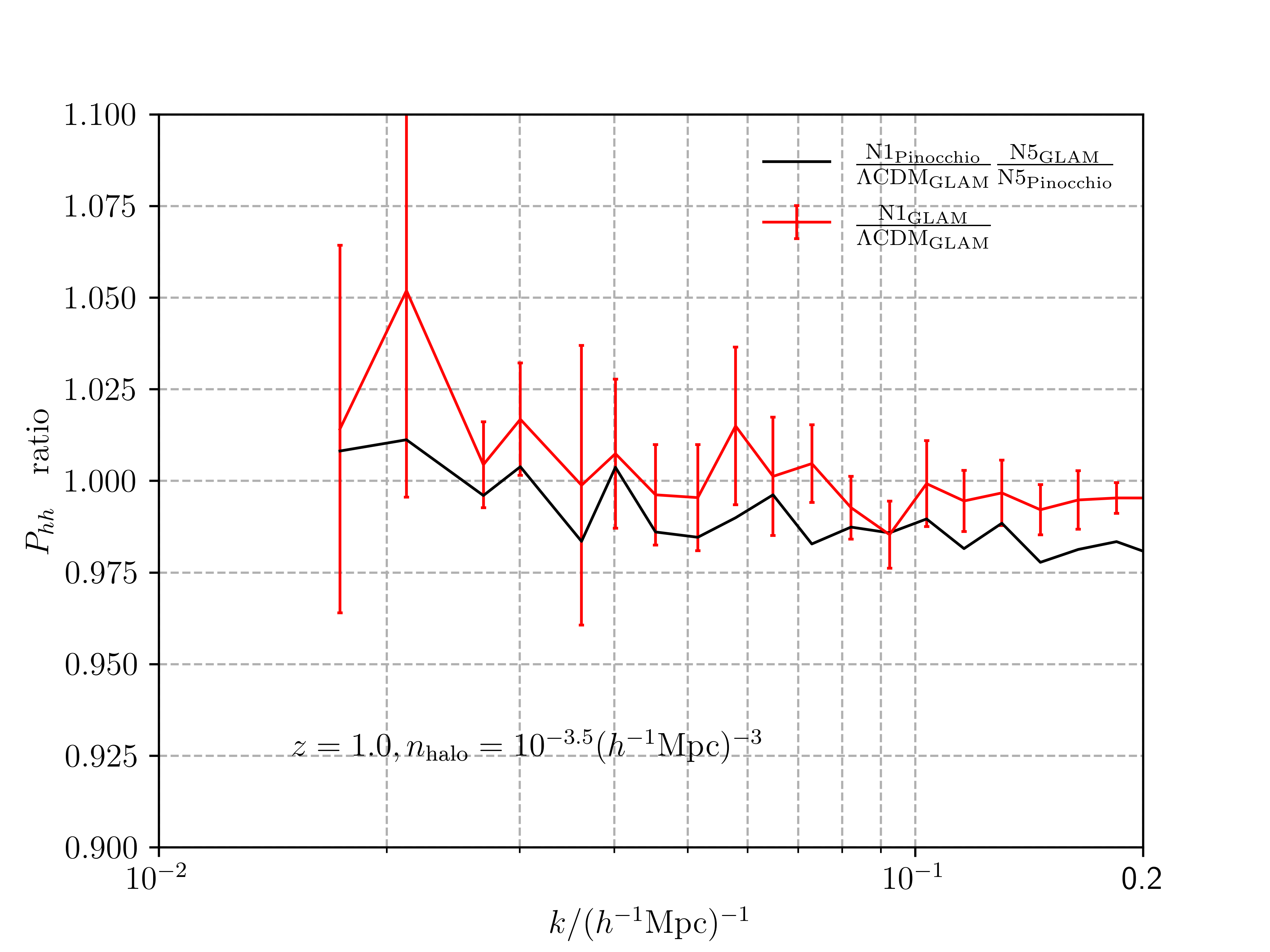} 
    \caption{Ratio of the halo power spectrum between N1 and $\Lambda$CDM. The red curves and error bars are from N-body simulations, while the black curves is the ratio between rescaled N1-PINOCCHIO and $\Lambda$CDM N-body. The upper and lower plots show the results for $z=0$ and $z=1$, respectively.}
    \label{fig:pkhh_res}
\end{figure}

In Fig.~\ref{fig:pkhh_res}, we show the ratio of halo power spectra between N1 and $\Lambda$CDM. The red curve with error bars are the results from N-body simulations, while the black curve is the averaged ratio between the re-scaled N1-PINOCCHIO and the $\Lambda$CDM N-body. The latter is within the scatter of the former. Furthermore,  one can see that the difference between N1 and $\Lambda$CDM models are about $2\%$. Compared with the effect in the matter power spectrum, eg. Figure 4 of \cite{Winther:2015}, the difference in the halo power spectrum is much smaller. This can be explained as follows: due to the enhancement of the gravitational force in nDGP, haloes which originated from less clustered lower initial density contrast peaks will now move to higher mass bins. As mentioned above, our catalogues are filtered to have the same number of halos both in nDGP and $\Lambda$CDM. Hence, the catalogs in nDGP have more massive halos and these halos trace the less clustered initial density field. This effect compensates the gravitational force enhancement, resulting in a halo power spectrum very similar to the $\Lambda$CDM one. A similar result has also been reported in the context of N-body simulations of $f(R)$ gravity \cite{Arnold:2019zup}.

\begin{figure}
	\centering
	\includegraphics[width=0.7\columnwidth]{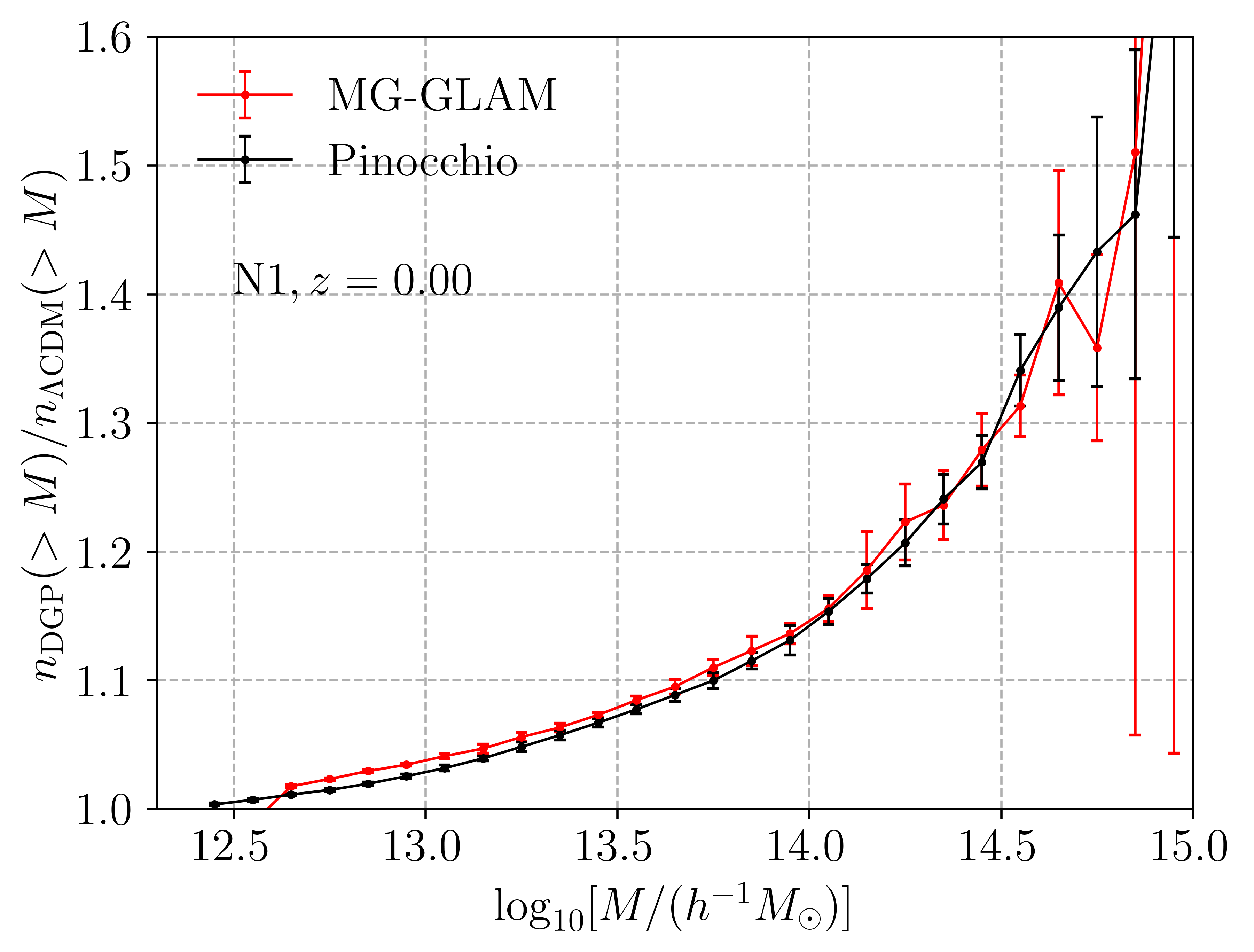}
	\includegraphics[width=0.7\columnwidth]{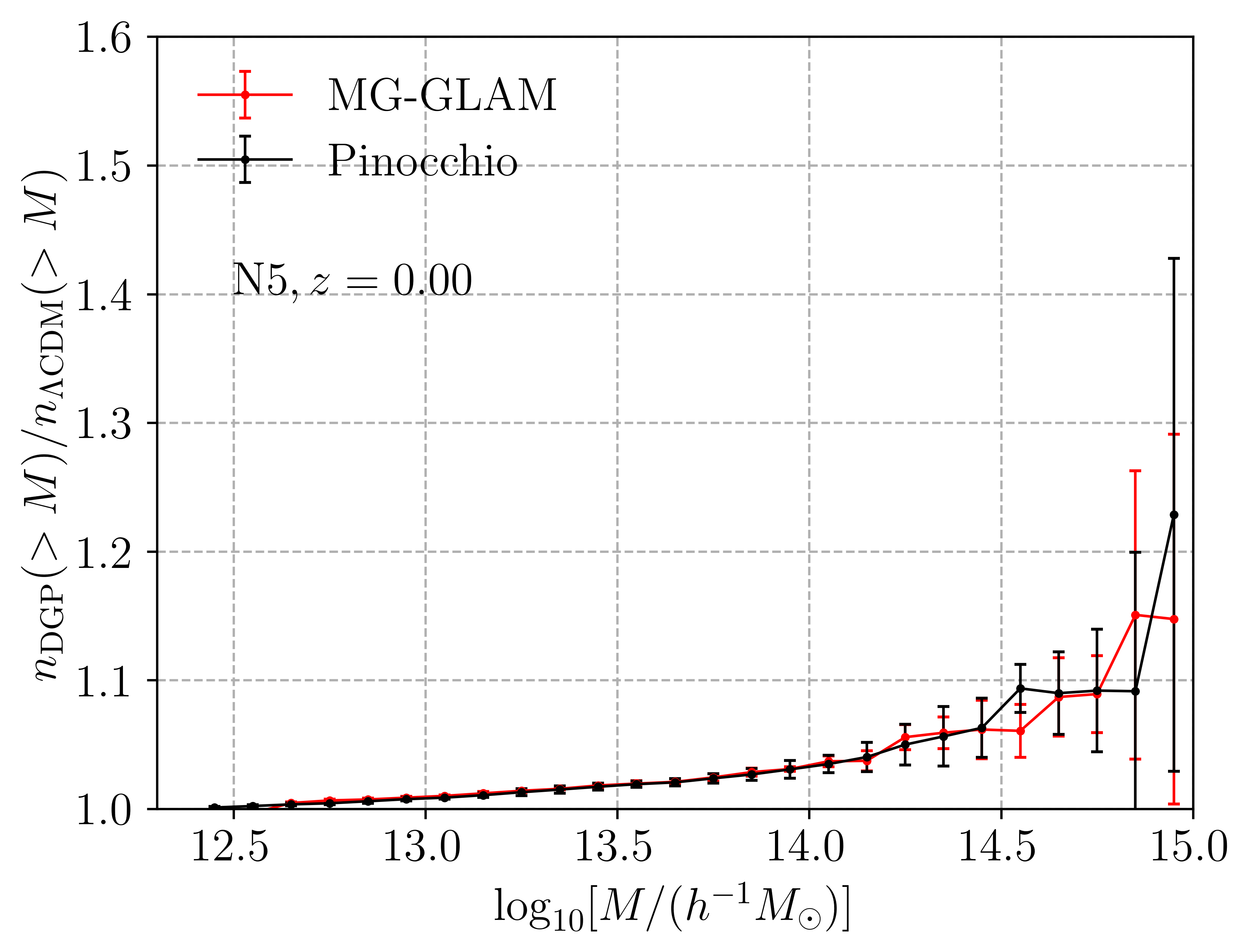}
    \caption{Ratio of the cumulative halo mass between nDGP and $\Lambda$CDM at $z=0$. The upper and lower plots show results for $r_cH_0=1$ and $r_cH_0=5$, respectively. Black and red squares with error bars are the results from PINOCCHIO and N-body simulation (MG-GLAM), respectively.}
    \label{fig:hmf0}
\end{figure}

In Fig. \ref{fig:hmf0}, we show the ratio of the cumulative halo mass function for nDGP with respect to $\Lambda$CDM. The red and black curves are from N-body and PINOCCHIO, respectively. The upper and lower plots show the $z=0$ results for N1 and N5, respectively. One can see that, except for the $M<10^{13}~h^{-1}M_{\odot}$ in the N1 case, all data points are in good agreement with the N-body simulation.

\section{Conclusions}\label{sec:6}

In this paper we study dark matter structure formation in the normal branch of the DGP model using the PINOCCHIO algorithm. For this purpose, we first present 2nd order Lagrangian perturbation theory for the nDGP model. The 1st- and 2nd-order growth functions in nDGP are larger than those in $\Lambda$CDM, which leads to stronger galaxy clustering in nDGP. Secondly we present the dynamics of ellipsoidal collapse in nDGP. Due to the enhancement of gravitational interaction, the collapsing process is accelerated in nDGP compared to $\Lambda$CDM. We create a new branch in the public PINOCCHIO repository for nDGP to store our implementation. The extended PINOCCHIO code provides a fast tool to generate dark matter particle snapshots and halo catalogues at given redshifts, as well as past light cones. To validate we run the nDGP-PINOCCHIO code in a box with size 512 ${\rm Mpc}~h^{-1}$ and 1024$^3$ particles and study the statistical properties of these catalogs, focusing on the real space halo power spectrum and halo mass function. As a showcase, we run the code with two values of the additional model parameter $r_c$, namely $r_c~H_0=1$ and $r_c~H_0=5$, respectively. We name the two N1 and N5 respectively, with the former corresponding to a larger modified gravity effect. We compare these results with those from N-body simulations. 

Due to the different definition of halo finders, there exists a systematic off-set in the halo power spectrum between N-body and PINOCCHIO. In order to mitigate this discrepancy, we calibrate the original N1-PINOCCHIO result by a ratio between N-body and PINOCCHIO of N5 model. The relative residuals are in the range of $5\%$ for the halo spectrum up to $k\sim 0.3$ at $z=0$, and $k\sim 0.5$ at $z=1$. Furthermore, we calculate the ratio between the halo power spectrum for N1 and $\Lambda$CDM, finding a difference of about $2\%$ between the two. Compared with the effect in the matter power spectrum, the difference in the halo power spectrum is much smaller, because the enhancement of the gravitational force in nDGP is compensated by the selection we perform on the halo catalogues to match number density across galaxy models. Hence, the catalogs in nDGP feature more massive halos and these halos trace the less clustered initial density field. As a result, the halo power spectrum in the calibrated N1 model has very little difference from the one in $\Lambda$CDM. 
For the cumulative halo mass function, the agreement is better in the higher mass end than the lower one. 

In order to explore the large parameter space for modified gravity and dark energy models, a fast machinery to generate large sets of halo catalogues is of paramount importance. This is needed for the construction of numerical covariance matrices for cosmological observables, which require a large number of realisations where observational systematics and survey-specific effects can be added. Additionally, novel inference methods such as simulation based inference require the generation of large sets of realisations. As a semi-analytical method, PINOCCHIO is very suitable for this task. In this paper, we take the nDGP model as a working example for scale-independent MG featuring Vainshtein screening. Our implementation is straight forward and formulated from first principles. The results match with N-body simulations from linear to mildly non-linear scales within a few percent accuracy, consistently with the performance of PINOCCHIO in $\Lambda$CDM. We believe we present a promising and computationally economic methodology for exploring the nature of dark energy.


\acknowledgments

YLS and BH are supported by the National Natural Science Foundation of China Grants No. 11973016. CM's work is supported by the Fondazione ICSC, Spoke 3 Astrophysics and Cosmos Observations, National Recovery and Resilience Plan (Piano Nazionale di Ripresa e Resilienza, PNRR) Project ID CN\_00000013 ``Italian Research Center on High-Performance Computing, Big Data and Quantum Computing'' funded by MUR Missione 4 Componente 2 Investimento 1.4: Potenziamento strutture di ricerca e creazione di ``campioni nazionali di R\&S (M4C2-19 )'' - Next Generation EU (NGEU). CM also acknowledges support from a UK Research and Innovation Future Leaders Fellowship [grant MR/S016066/2] for the early stages of this project.


\bibliographystyle{unsrt}
\bibliography{example}



\end{document}